\documentclass[final,5p,times,twocolumn,numbers]{elsarticle}

\usepackage{epsfig}
\usepackage[utf8]{inputenc} 
\usepackage[T1]{fontenc}    
\usepackage{url}            
\usepackage{booktabs}       
\usepackage{amsfonts}       
\usepackage{amsmath}        
\usepackage{amssymb}        
\usepackage{amsthm}         
\usepackage{nicefrac}       
\usepackage{microtype}      
\usepackage{xcolor}         
\usepackage{verbatim}       
\usepackage{graphicx}       
\usepackage{hyperref}       
\usepackage[ruled,noend]{algorithm2e}
\usepackage[capitalise,noabbrev]{cleveref}

\crefname{equation}{}{}
\usepackage[ruled,noend]{algorithm2e}

\theoremstyle{plain} 

\theoremstyle{definition} 

\theoremstyle{remark} 

\newcommand{\E}{\mathbb{E}} 
\newcommand{\R}{\mathbb{R}} 
\newcommand{\N}{\mathbb{N}} 

\journal{International Journal of Forecasting (IJF)}

\begin{document}

\begin{frontmatter}



\title{An adaptive volatility method for probabilistic forecasting and its application \\ to the M6 financial forecasting competition}


\author[su,viking]{Joseph de Vilmarest\corref{contrib}}
\ead{joseph.de-vilmarest@vikingconseil.fr}
\ead[url]{https://vikingconseil.fr/}

\author[su,sdu]{Nicklas Werge\corref{contrib}}
\ead{werge@sdu.dk}
\ead[url]{https://nicklaswerge.github.io}

\affiliation[su]{organization={Sorbonne Université},
            city={Paris},
            country={France}}

\affiliation[viking]{organization={Viking Conseil},
            city={Paris},
            country={France}}
            
\affiliation[sdu]{organization={University of Southern Denmark},
            city={Odense},
            country={Denmark}}

\cortext[contrib]{Authors contributed equally}


\begin{abstract}
In this paper, we address the problem of probabilistic forecasting using an adaptive volatility method rooted in classical time-varying volatility models and leveraging online stochastic optimization algorithms. These principles were successfully applied in the M6 forecasting competition under the team named {\it AdaGaussMC}. Our approach takes a unique path by embracing the Efficient Market Hypothesis (EMH) instead of trying to beat the market directly. We focus on evaluating the efficient market, emphasizing the importance of online forecasting in adapting to the dynamic nature of financial markets. The three key points of our approach are: (a) apply the univariate time-varying volatility model AdaVol \citep{werge2022adavol}, (b) obtain probabilistic forecasts of future returns, and (c) optimize the competition metrics using stochastic gradient-based algorithms. We contend that the simplicity of our approach contributes to its robustness and consistency. Remarkably, our performance in the M6 competition resulted in an overall 7\textsuperscript{th} ranking, with a noteworthy 5\textsuperscript{th} position in the forecasting task. This achievement, considering the perceived simplicity of our approach, underscores the efficacy of our adaptive volatility method in the realm of probabilistic forecasting.
\end{abstract}

\begin{keyword}
forecasting competition \sep financial forecasting \sep probabilistic forecasting \sep efficient market hypothesis \sep online learning \sep volatility

\MSC 91B28 \sep 91B84
\end{keyword}

\end{frontmatter}

\section{Introduction} \label{sec:introduction}
In financial time-series analysis, the practice of time-series forecasting stands as an indispensable element \citep{brockwell2002introduction,box2015time,hamilton2020time,hyndman2018forecasting}. The ability to accurately predict future time-series is essential for decision-makers in the complex landscape of financial markets. Building upon the legacy of the five preceding M competitions, each dedicated to advancing methods in time-series forecasting \citep{makridakis1982accuracy,makridakis1993m2,makridakis2000m3,makridakis2020m4,makridakis2022m5,makridakis2022m5acc}, the M6 competition emerges as a pivotal chapter in the exploration of forecasting methodologies \citep{makridakis2023m6}. With a specific focus on scrutinizing the tenets of the Efficient Market Hypothesis (EMH), the M6 competition aims to provide valuable insights for researchers and practitioners interested in exploring the relationship between probabilistic forecasting and investment decision-making. Its primary objective is to bring fresh insights to the forefront of the EMH, a hypothesis positing that share prices encapsulate all relevant information. This notion implies that consistently outperforming the market is not feasible.

The M6 competition was composed of two parts: \emph{probabilistic forecasting} and \emph{investment decision-making}. The participants underwent a rigorous live evaluation process that occurred monthly for a duration of twelve months. In the forecasting part, the goal was to predict the rank probability of each financial asset, focusing on returns in the 1\textsuperscript{st} to 5\textsuperscript{th} quantile. The investment part involved deciding whether to invest or not based on the forecasted probabilities. These component added complexity, as participants had to not only decide on individual investments but also construct portfolios that matched their forecasts. The challenge extended to a diverse investment universe comprising $50$ S\&P500 stocks and $50$ international Exchange Traded Funds (ETFs), spanning various asset categories and countries.

Instead of challenging the EMH, our approach was to embrace it, particularly under a non-stationary market paradigm. In the M6 competition, our strategy had a twofold focus: i) embracing the principles of an efficient market and ii) adapting to the dynamics of a non-stationary market paradigm.

For the EMH, we asked a fundamental question:
\begin{center}
What would an efficient market do?
\end{center}
Our methodology exclusively employed time-series methods, specifically focusing on the daily returns of assets and avoiding the inclusion of any external data. By estimating the future distribution of returns based on historical data, our goal was to evaluate the efficient market's probabilistic forecast.

Concentrating on the principle that
\begin{center}
an efficient market implies that the expected return should be the same for a given level of risk,
\end{center}
we directed our efforts towards modeling volatility. Forecasting volatility from historical data was key to assessing the efficient market's perception of the risk associated with each asset. This approach formed the foundation for our probabilistic forecast submission to the M6 competition.

To adapt to the challenges of a non-stationary market paradigm, we employed online learning \citep{cesa2006prediction}. This methodology enabled us to dynamically adjust our model in response to evolving market conditions, ensuring the robustness of our forecasting methodology in the face of changing trends and uncertainties. The integration of efficient market principles with an adaptive response to market dynamics positioned our approach as a versatile and responsive method for probabilistic forecasting in financial time-series analysis.

While there are many ways to evaluate volatility, we advocate for online learning methods \citep{cesa2006prediction}. This paradigm has been applied to various fields and performed particularly well in recent electricity load forecasting competitions \citep{gaillard2016additive,de2022state}.
The strength of this adaptive approach is to take account for regime-changes in data, i.e., non-stationarity. Applied to volatility \citep{werge2022adavol}, it allows to account for temporary breaks in the data with periods of very high-volatility, such as the recent COVID crisis. Our hypothesis is that 
\begin{center}
online learning procedures are necessary for forecasting future returns in an efficient market, given its non-stationary nature.
\end{center}

This paper provides an in-depth discussion of our online methodological approach, which secured a 5\textsuperscript{th} rank in the forecasting task and a 7\textsuperscript{th} rank in the M6 competition overall. Therefore, it serves as a comprehensive discussion paper outlining our approach and its results in the dual forecasting challenges. Notably, our strategy consistently surpassed the naive benchmark in probabilistic forecasting, showcasing the effectiveness of employing online volatility models, particularly AdaVol \citep{werge2022adavol}. We emphasize the frugality of our methods, asserting that simplicity contributes to the robustness and consistency observed in our results.

{\bf Organization.} In \cref{sec:lowrisk} we introduce our approach to the M6 competition with simple methods beating the naive benchmark for the forecasting task. In \cref{sec:adavol}, we present the adaptive volatility model, AdaVol. This is followed by the application of AdaVol to the M6 competition in \cref{sec:m6}. \Cref{sec:discussion} contains a discussion of the performance of our methodology in the M6 competition.

\section{Beating the M6 Forecasting Benchmark with Low Risk} \label{sec:lowrisk}

The M6 financial forecasting competition aimed at assessing the EMH. To that end, it considered a financial universe composed of 100 assets: half stocks and half ETFs. The competition consisted in two tasks:  \emph{probabilistic forecasting} where the objective was to rank the future returns of the 100 assets in a probabilistic fashion, and \emph{investment decision-making} where the competitors proposed portfolio allocations. Each task had a 4-week horizon, and there were 12 predictions to make, so the competition lasted roughly a year.

Our intuition is that it is extremely hard to beat the market. However, we remarked that the benchmark proposed by the competition organizers was not really {\it following the market}, and therefore we understood that there were simple ways to outperform this benchmark.

We focus on the \emph{probabilistic forecasting} task. The objective was to rank the 100 assets in probability. More specifically, at each point the assets were put in five quantiles of 20 with respect to their 4-week returns. Participants were asked to assess the probability of each asset falling in each quantile, that is 500 (discrete) probabilities. The naive benchmark was uniform, that is $20\%$ for each asset and each quantile. However, the EMH does not say that the distribution of the future return of each financial asset should be the same. The EMH rather proposes that the expected return of each financial asset for a given level of risk should be the same. In other words, some assets are more volatile than others, and for these risky assets a higher return is expected.

In the competition the universe was composed of 50 stocks and 50 ETFs. Generally, stocks are much more volatile than ETFs which are diversified. Therefore, attributing a probability higher than 20\% for ETFs befalling in the middle quantile, and lower than 20\% for these assets befalling in the extreme quantiles, seems natural. More precisely, we display the historical quintiles in \cref{tab:hist_quintile}. We classify the assets in five classes: Stocks, ETF Equities, ETF Fixed income, ETF Commodities and ETF Volatility. Then, we compute the frequency of each quintile for each class. For instance, during the considered period, stocks appeared in the extreme quintiles $28.7\%$ and $23.7\%$ of the time, while only $14.4\%$ of the instances in the middle one.
\begin{table*}[th]
    \centering
    \begin{tabular}{c|c|c|c|c|c|c}
        Class & Asset IDs & Q1 & Q2 & Q3 & Q4 & Q5 \\
        \hline
        Stocks & 1-50 & 28.7\% & 18.1\% & 14.4\% & 15.1\% & 23.7\% \\
        ETF Equities & 51-67, 80-99 & 10.9\% & 24.2\% & 27.8\% & 23.9\% & 13.1\% \\
        ETF Fixed income & 68-76 & 9.5\% & 15.1\% & 23.2\% & 34.6\% & 17.6\% \\
        ETF Commodities & 77-79 & 21.0\% & 14.6\% & 14.5\% & 17.5\% & 32.4\% \\
        ETF Volatility & 100 & 28.7\% & 3.7\% & 1.8\% & 3.2\% & 62.5\%
    \end{tabular}
    \caption{Frequency of each quintile for each class during years 2015 to 2020. Quintile Q1 is composed of the 20 assets performing the best, while Q5 is composed of the 20 assets performing the worse.}
    \label{tab:hist_quintile}
\end{table*}

We propose two very simple benchmarks to motivate our approach, and the gain obtained with respect to the competition one provides an explanation on our performance.

As presented above, the \emph{probabilistic forecasting} task at time $t$ consists in submitting a matrix of $100\times 5$ entries, that we denote by $M_t$. The loss function used by the competition is the Ranked Probability Score (RPS) between the submission and the true value, that we denote by $Q_t$. Our first observation is that the historical values of $Q_t$ are not uniform in $[0,1]$; see Table \ref{tab:hist_quintile}. Therefore, a very simple idea is to use the best constant matrix $M$ with respect to the historical RPS. For some period $\mathcal{T}\in\N$, we compute
\begin{align*}
    M_{\mathcal{T}} \in \arg\min\limits_{M\in[0,1]^{5\times 500}} \sum\limits_{t\in\mathcal{T}} RPS(M,Q_t) \,.
\end{align*}
We observe that $M$ is equivalently defined by
\begin{align*}
    M_{\mathcal{T}} = \frac{1}{|\mathcal{T}|} \sum\limits_{t\in\mathcal{T}} Q_t\,.
\end{align*}
Indeed, the RPS is a quadratic loss between linear transforms of $M$ and $Q_t$.

Therefore, a simple benchmark, named {\it best constant}, consists in defining a training set $\mathcal{T}$ (for instance, years 2015 to 2020) and submitting constantly $M_{\mathcal{T}}$.

This very simple method is purely based on the competition metrics. However, our fundamental goal is probabilistic time-series forecasting. We aim to forecast the log-return of each asset in the 4-week horizon (a matrix $R_t\in\mathbb{R}^{20\times 100}$ for any instant $t$). Our first approach is to predict the marginal distributions of each component and assume their independence. We fit Gaussian distributions for each marginal on a training set $\mathcal{T}$: the daily log-return of asset $a$ is predicted with $r_t\sim\mathcal{N}(\hat\mu_a, \hat\sigma_a^2)$. Relying on the EMH, we don't claim to be able to predict $\hat\mu_a$ better than the market does. We assume that for a given risk the expected return should be the same. We do the approximation that for a \emph{class} of financial asset the risk is the same, and therefore the expected returns are similar. We set four classes: stocks, ETF equities, ETF fixed income, ETF commodities; therefore, $\hat{\mu}_a$ is the average return of all assets of the corresponding class during years 2015 to 2020.

This yields a probabilistic forecast $R_t\sim P_t$. Based on that forecast, we minimize the expected RPS as follows:
\begin{align*}
    M_t\in \arg\min\limits_{M\in[0,1]^{5\times 500}} \mathbb{E}_{R_t\sim P_t} [RPS(M,Q_t)] \,,
\end{align*}
with $Q_t$ defined by the ranks of $R_t$. The RPS is convex, therefore we minimize it simply with a stochastic gradient descent of annealing step size, under Monte-Carlo simulation of $R_t$, see Section \ref{sec:opt}.

\begin{figure}[th]
    \centering
    \includegraphics[width=7cm]{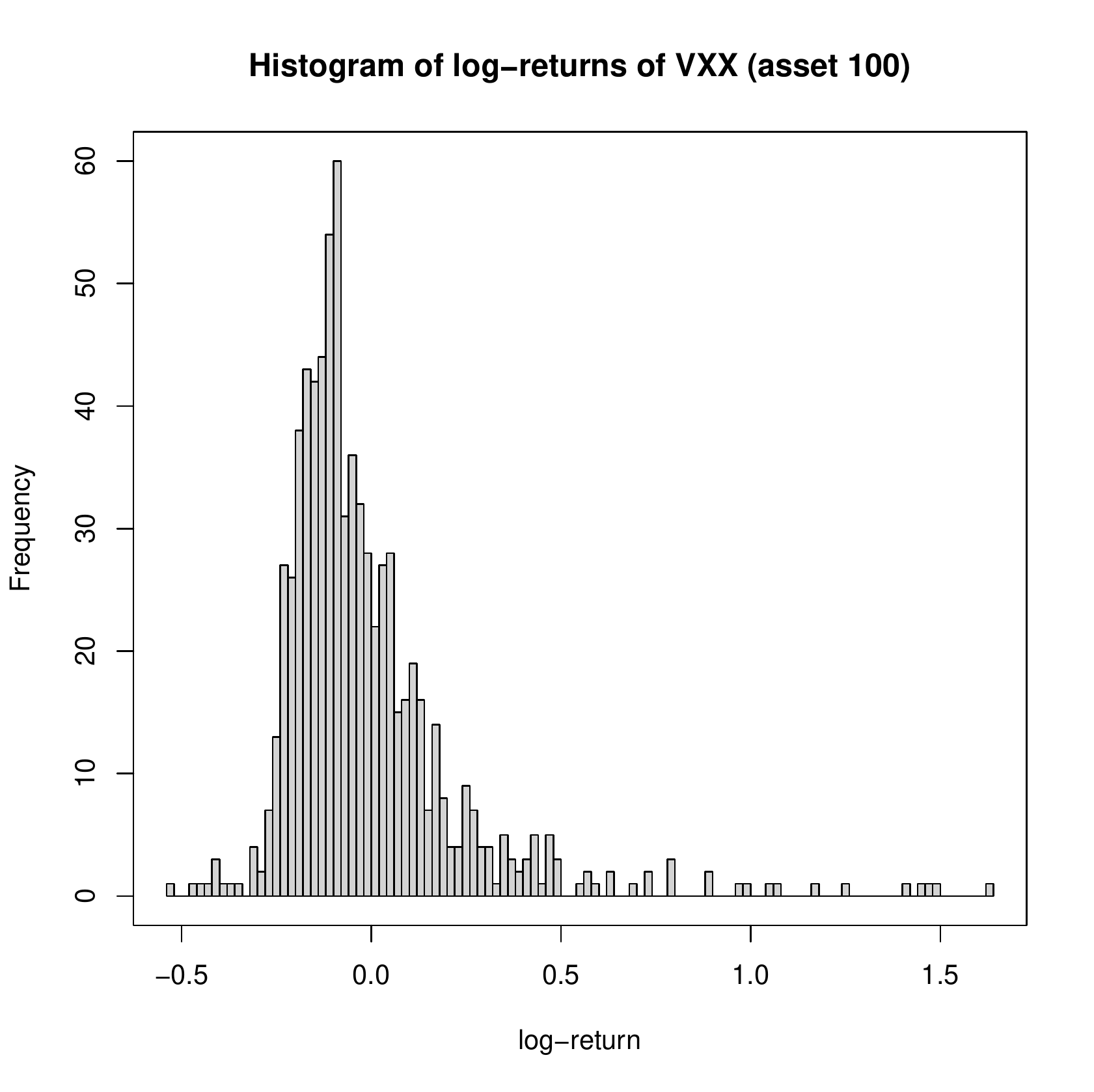}
    \caption{Histogram of VXX log-return during years 2015 to 2020. During that period, this asset falls essentially in extreme quintiles (62.5\% in the worse, 28.7\% in the best), see Table \ref{tab:hist_quintile}.}
    \label{fig:return_vxx}
\end{figure}

We evaluate on the last 9 months of 2021; that is the data available before the start of the competition and without missing value for the assets. The simple {\it best constant} benchmark achieves a RPS of 0.1570, while the naive M6 benchmark has a RPS of 0.16. Then, our optimization of the expected RPS based on a Gaussian distribution yields a very close RPS of 0.1571. Finally, we observe that the log-returns of the final asset (volatility ETF) are far from Gaussian ({\it c.f.} Figure \ref{fig:return_vxx}); more importantly, these log-returns are strongly asymmetric, which makes it sub-optimal to use a symmetric distribution. Therefore, we define a hybrid forecast keeping our Gaussian distribution for the first 99 assets and drawing the last uniformly from its past values (2015 to 2020). This achieves a RPS of 0.1567. The difference between these benchmarks seems very small; however, it is significant as one should note that the differences of performance between top teams in the competition were similar. There is a bigger gap between the competition naive benchmark and {\it best constant}, than between {\it best constant} and the top-performing team.

From these simple benchmarks, we draw two conclusions: First, while it is hard to beat the market, it is easy to beat the uniform benchmark because the different assets are associated to different risks; assets with high volatility (stocks) are more likely to fall in the extreme quantiles than assets with low volatility (ETFs). Second, excluding the last asset (volatility ETF), it is possible to model each asset's distribution with a Gaussian, where we don't focus on forecasting the mean but on the variance; indeed, we apply a simple assumption on the mean (assets of a same class have the same expected return), but we aim to capture the different behaviors with volatility forecasting.

Our Gaussian forecasting model seems very naive. Nonetheless, its advantage is to yield a framework on which we can apply more sophisticated methods. Indeed, we rely on online learning to estimate the volatility. The objective is to capture the evolution of each asset's volatility to enhance our probabilistic forecast. That is presented in the next section.

\section{AdaVol: an Adaptive Volatility Method} \label{sec:adavol}

The intricate nature of financial time-series reveals a dynamic characteristic in volatility, characterized by its time-varying nature and frequent clustering phenomena. The quest to model and predict this volatility has led to the exploration of various methodologies, with non-linear time-series models often taking center stage. Among these, the AutoRegressive Conditional Heteroskedasticity (ARCH) model and the Generalized ARCH (GARCH) model are the most well-known \citep{engle1982autoregressive,bollerslev1986generalized}.

However, the GARCH model hinges on the assumption of stationarity, a premise that might be subject to scrutiny in real-world financial data. The inherent non-stationarity of financial time-series data prompts the exploration of alternative approaches, with a natural inclination towards adaptive methods for robust volatility modulation. To solve this, we consider AdaVol \citep{werge2022adavol}, an innovative online volatility method designed to navigate the challenges posed by time-varying volatility in financial data.

AdaVol departs from traditional stationary assumptions and embraces adaptability as a cornerstone for modeling volatility dynamics. By leveraging the principles of online learning, AdaVol addresses the limitations of GARCH and offers a flexible framework to capture the nuanced evolution of volatility over time. This departure from stationarity assumptions aligns AdaVol with the inherent characteristics of financial time-series data, where volatility is known to evolve dynamically. Specifically, AdaVol's adaptability during regime-changes, as evidenced in  \citet[Figure 8]{werge2022adavol}, and its capacity to react to major events like the COVID-19 pandemic, as illustrated in \citet[Figure 9]{werge2022adavol}, highlight its capability to address the complexities of time-varying volatility in financial data.

In its simplest form, AdaVol is a GARCH-like model, where the statistical inference is carried out using the Quasi-Maximum Likelihood (QML) procedure, which is recursively updated using stochastic gradient-based algorithms \citep{bottou2018optimization}. This methodology enables AdaVol to recursively update its estimates based on incoming data, ensuring a responsive and dynamic adaptation to changing volatility patterns. Unlike GARCH, AdaVol is more well-suited for the inherent non-stationarity observed in financial time-series.

As outlined in \citet{werge2022adavol}, AdaVol is constructed from the centered GARCH$(p,q)$ process $(\epsilon_{t})$ with variance targeting estimation \citep{francq2011merits}; 
\begin{align*}
    \begin{cases}
    \epsilon_{t} = \sigma_{t} \eta_{t},\\
    (\sigma_{t}^{2} - \gamma_{t-1}^{2}) =  \sum_{i=1}^{p} \alpha_{i}  (\epsilon_{t-i}^{2}-\gamma_{t-1}^{2}) + \sum_{j=1}^{q} \beta_{j} (\sigma_{t-j}^{2}-\gamma_{t-1}^{2}),
    \end{cases}
\end{align*}
where $\{\alpha_{i}\}_{i=1}^{p}$ and $\{\beta_{j}\}_{j=1}^{q}$ are non-negative parameters ensuring the non-negativity of the variance process $(\sigma_t^2)$, $(\eta_t)$ is a sequence of i.i.d. random variables with $\E[\eta_{0}]=0$ and $\E[\eta_{0}^{2}]=1$, and $(\gamma_{t})$ is the sample volatility of $(\epsilon_{i})_{i<t}$.

The usual approach for estimating parameters $\theta=(\alpha_{1},\dots,\alpha_{p},\beta_{1},\dots,\beta_{q})^{\top}\in\R_{+}^{p+q}$ is by the QML estimator \citep{berkes2003,francq2004,straumannmikosch2007}. Here, the goal is to minimize the Quasi-Likelihood function $L_{n}(\theta)$ defined by
\begin{align*}
L_{n}(\theta) = \sum_{t=1}^{n}l_{t}(\theta),
\end{align*}
with QL loss
\begin{align*}
l_{t}(\theta)=\log(\sigma_{t}(\theta))+X_{t}^{2}/\sigma_{t}^{2}(\theta).
\end{align*}
Remark, these parameters $\theta$ are used in the volatility process $\sigma_{t}^{2}(\theta)$ to make volatility forecast.

Commonly, iterative estimation procedures are used for the minimization of $L_{n}(\theta)$, e.g., quasi-Newton methods \citep{nocedal1999numerical}.
Roughly, each iteration will have a computational cost of $\mathcal{O}(n(p+q))$, making the minimization cost $\mathcal{O}(nm(p+q))$, where $m$ is the number of iterations. As new data arrive, this becomes prohibitively expensive and increasingly computationally inefficient. Furthermore, iterative optimization tools are unsuitable for financial data, as data often arrives in large quantities and with high frequency.

Stochastic optimization procedures are undoubtedly advantageous since observations are processed one-by-one \citep{bottou2018optimization}; this is very scalable as the cost is only $\mathcal{O}(p+q)$ computations for the minimization (compared to $\mathcal{O}(nm(p+q))$).
In online QML estimation, the parameter estimate is updated exclusively based on the previous estimate and the new observation. Thus, computationally efficient, as each new observation only need to processed once.

For AdaVol, our optimization strategy leverages first-order stochastic optimization methods, employing AdaGrad as the learning rate \citep{duchi2011adaptive}. To ensure adherence to the parameter space constraints, we augment this approach with a projected version. This combination not only enhances the convergence speed of the optimization process but also guarantees that the estimated parameters remain within the valid parameter space, reinforcing the stability and reliability of AdaVol's volatility forecasts. Specifically, AdaVol minimizes $L_{n}(\theta)$ by $\theta_{n}$, which is derived from the recursion:
\begin{align*}
\theta_{t} = \mathcal{P}_{\Theta} \left[\theta_{t-1}-\frac{\eta}{\sqrt{\sum_{i=1}^{t} \nabla_{\theta}l_{i}(\theta_{i-1})^{2}+\epsilon}} \nabla_{\theta}l_{t} (\theta_{t-1}) \right], \quad \theta_{0}\in\Theta,
\end{align*}
where $\eta>0$ is a constant learning rate, $\epsilon>0$ a small number ensuring positivity of the denominator, and $\mathcal{P}_{\Theta}$ is the projection onto
\begin{equation*}
\Theta=\left\{\left(\alpha_1,\ldots,\alpha_p,\beta_1,\ldots,\beta_q\right)\in\R_+^{p+q} \middle| \sum_{i=1}^p \alpha_i  + \sum_{j=1}^q \beta_j < 1 \right\}.
\end{equation*}

\Cref{algo:adavol} describes AdaVol in more detail. Note $\nabla_{\theta}l_{i}(\theta_{i-1})^{2}$ denotes the element-wise square $\nabla_{\theta}l_{i}(\theta_{i-1})\odot\nabla_{\theta}l_{i} (\theta_{i-1})$. Additionally, a practical implementation of AdaVol can be explored on GitHub.\footnote{\url{https://github.com/nicklaswerge/AdaVol}}

\begin{algorithm}[tp]
\caption{AdaVol \citep{werge2022adavol}} \label{algo:adavol}
\SetAlgoLined
\DontPrintSemicolon
\SetKwComment{Comment}{/* }{ */}
\KwIn{$\theta_{0}\in\Theta\subseteq\R^{p+q}$, $\eta=0.1$, $\epsilon=10^{-8}$}
\KwOut{$\sigma_{t+1}^{2}$ (forecasted volatility)}
Initialization: $\sigma_{1}^{2} = \epsilon_{1}^{2}$, $\mu_{0}=0$, $\gamma_{0}^{2} = 0$, $G_{0} = \epsilon$ and $t=0$\;
\For{each data sample $\epsilon_{t}$}{
$t=t+1$\Comment*[r]{time}
$\mu_{t}=t(t+1)^{-1}\mu_{t-1}+(t+1)^{-1}\epsilon_{t}$\Comment*[r]{mean}
$\gamma_{t}^{2}=(t-1)t^{-1}\gamma_{t-1}^{2}+t^{-1} (\epsilon_{t}-\mu_{t})^{2}$\Comment*[r]{variance}
$g_{t}=\nabla_{\theta}l_{t}(\theta_{t-1})$\Comment*[r]{gradient}
$G_{t}=G_{t-1}+g_{t}^{2}$\Comment*[r]{squared gradients}
$\theta_{t}=\mathcal{P}_{\Theta}[\theta_{t-1}-\eta G_{t}^{-1/2}g_{t}]$\Comment*[r]{estimates}
$\sigma_{t+1}^{2}(\theta_{t})$\Comment*[r]{forecast volatility}
}
\end{algorithm}

AdaVol's architecture has demonstrated its efficacy in generating robust and adaptable forecasts. Its capability to adjust to time-varying parameters proves advantageous in scenarios characterized by non-stationarity. Additionally, AdaVol stands out for its computational and memory efficiency, leveraging only the preceding (GARCH) estimate to process new observations. This streamlined approach ensures a single pass through the observations, minimizing computational overhead.

In \citet[Appendix B]{werge2022adavol}, the authors conducted a relative computational speed comparison, demonstrating that AdaVol is approximately $205$ times faster than the GARCH$(1,1)$ model for a sample size of $n=1000$ \citep[Table B.4]{werge2022adavol}. Furthermore, they observed that the relative speed gain of AdaVol improves with larger sample sizes.

It is noteworthy that financial data commonly arrives in time-varying mini-batches. A straightforward extension of AdaVol to this dynamic setting results in a computational cost of just $\mathcal{O}(b_{t}(p+q))$, where $b_{t}$ denotes the number of observations arriving at time $t$. This approach aligns with the frequent occurrence of time-varying mini-batches in financial data. Simultaneously, the adoption of time-varying mini-batches has been substantiated to enhance the estimation procedure \citep{godichon2023non,godichon2023learning}.

\section{Back to M6 Financial Forecasting Competition}\label{sec:m6}

As presented in Section \ref{sec:lowrisk}, the M6 financial forecasting competition aimed at assessing the EMH. It consisted in two tasks: \emph{probabilistic forecasting} and \emph{investment decision-making}.

We did not use specific knowledge on finance; our strategy was based solely on probabilistic time-series forecasting and stochastic optimization.
Indeed, each task was evaluated by a metrics and our objective was to optimize it.

We developed a strategy in two steps. First, as our aim is probabilistic time-series forecasting, we obtain such probabilistic forecasts using AdaVol. Then, our second step consists in optimizing the expected loss function with respect to the submission based on these probabilistic forecasts.

\subsection{Probabilistic forecasting based on AdaVol}
Our objective is to forecast the return of each asset in the 4-week horizon (a matrix $R_t\in\mathbb{R}^{20\times 100}$). We first predict the marginal distributions of each component; then we reconcile them either with an independent assumption or after the estimation of correlations.

The prediction of the marginals fits in the setting of univariate time-series forecasting.
We denote by $r_{t,a}$ the log-return of asset $a$ at time $t$. We apply AdaVol on $(r_{t,a}-\hat\mu_a)_t$ assumed independent, where $\hat{\mu}_a$ is the estimated mean return per class defined in Section \ref{sec:lowrisk}. The Gaussian application of AdaVol yields $r_{t,a}\sim\mathcal{N}(\hat{\mu}_a,\hat\sigma_{t,a}^2)$. At submission point $i$ of time $t_i$, we have a volatility $\hat\sigma_{t_i,a}$, and our model becomes fixed:
\begin{align*}
    r_{t,a}\sim\mathcal{N}(\hat{\mu}_a,\hat\sigma_{t_i,a}^2),\quad t\ge t_i\,.
\end{align*}

We treat separately the final asset (volatility): we simply use the empirical distribution of its past returns.

Finally, we combine the marginals to obtain a distribution on $R_{t_i}$. We compared different approach on the year of data preceding the competition. The best results were not the same for the two tasks. For \emph{probabilistic forecasting}, we simply use independent assets, and our joint distribution is the product of marginals; for \emph{investment decision-making}, we estimated correlations between asset returns $(r_{a,a'})$ and our joint distribution was a multivariate Gaussian distribution of covariance matrix
\begin{align*}
    \begin{pmatrix}
    \hat\sigma_{t_i,1}^2 & \hat\sigma_{t_i,1}\hat\sigma_{t_i,2}r_{1,2} & \hdots \\
    \hat\sigma_{t_i,2}\hat\sigma_{t_i,1}r_{2,1} & \hat\sigma_{t_i,2}^2 & \hdots \\
    \vdots & \vdots & \ddots
    \end{pmatrix} \,.
\end{align*}

\subsection{Optimization of the expected loss function}\label{sec:opt}
For each task, our submission is a vector $x_{t_i}\in\mathbb{R}^p$; we have a loss function (the negative information ratio is minimized), denoted by $\ell$, that depends on our prediction and the return matrix $R_{t_i}$: the evaluation is $\ell(x_{t_i},R_{t_i})$.

The evaluation is the average of the loss on 12 iterations. 
As we have a probabilistic prediction of $R_i$, it is natural to minimize the expected loss obtained under that distribution. If our distribution was correct this would be optimal for a very large number of submission points. Our objective is the following:
\begin{align*}
    x_{t_i}\in\arg\min \mathbb{E}_{R_{t_i}\sim P_{t_i}} [\ell(x_{t_i},R_{t_i})] \,.
\end{align*}
Our optimization procedure relies on ADAM \citep{kingma2015adam}. Each optimization step relies on a mini-batch of $100$ samples of $P_{t_i}$, and the gradient step used is $\alpha_k=\mathcal{O}(1/\sqrt{k})$, where $k$ is the iteration number.

The convex nature of the RPS yields convergence of this procedure to the optimal point for \emph{probabilistic forecasting}, under the assumption that our distribution $P_{t_i}$ is correct. The negative information ratio is not convex and there is no guarantee of convergence to the optimal point for \emph{investment decision-making}. As a sanity check, we ensure that the attained point $x_{t_i}$ yields a better expected information ratio than the naive uniform portfolio allocation. During the competition we observed that our obtained information ratio was slightly above the one of the uniform benchmark, confirming this property, but not significantly better.

\begin{table}
    \centering
    \begin{tabular}{c|c|c|c|c|c}
        Task & $p$ & Assets & $\ell$ & Rank & Naive \\
        \hline
        Forecasts & $500$ & independent & convex & 5 & 39 \\
        Decisions & $100$ & correlated & not convex & 42 & 48
    \end{tabular}
    \caption{Summary of the specificities of each task.}
    \label{tab:tasks}
\end{table}

\section{Discussion}\label{sec:discussion}
Our objective was on the \emph{probabilistic forecasting} part of the M6 competition; here, we were ranked 5\textsuperscript{th} out of 163 competitors. Note that only 38 participants outperformed a naive benchmark designed by the organizers ({\it M6 dummy}). This achievement underscores the potential of our online approach to contribute significantly to understanding market dynamics and intricacies in financial time-series analysis.

In contrast, our performance in the \emph{investment decision} task was not as prominent, securing a 42\textsuperscript{nd} position compared to the benchmark's 48\textsuperscript{th} position. 

Overall, we ranked 7\textsuperscript{th} in the M6 competition. It has been established by the previous M competitions that \emph{statistically sophisticated methods do not necessarily outperform simple methods} \citep{makridakis2000m3}. We claim that the simple complexity of our methodology explains its robustness and this good performance.

\subsection{Interpretation of our Results}
These observations prompts three key interpretations that shed light on our strategy's behavior in the two tasks.

\paragraph{Market understanding}
The first interpretation is philosophical. As stated in the introduction, we did not try to contradict the EMH. Instead, we embraced it, and asked ourselves what would the probabilistic forecast of the efficient market be. Our in-depth analysis of the RPS in Section \ref{sec:lowrisk} explains how the uniform benchmark can be outperformed using probabilistic forecast of asset returns. Therefore, it is natural to obtain good performances in the probabilistic forecasting task. We did not conduct such study of the information ratio, and we conjecture that those who outperformed us possess a deeper understanding of the market. This acknowledgment emphasizes the complexity of financial markets and encourages continuous exploration and refinement of our approach to align with the dynamic and nuanced nature of market behavior.

\paragraph{Univariate vs. multivariate forecasting}
A second explanation lies in the distribution representation. Our strategy rely on univariate time-series forecasting from AdaVol, emphasizing correlations for the investment decision task while employing the product of marginals from the probabilistic forecasting task. The performance metric, Ranked Probability Score (RPS), favors an individualized approach as it is the sum of each asset's RPS. Correctly predicting the outcome for one asset minimally impacts the RPS of another. Hence, it would be possible to forecast independently the relative performance of each asset.

However, for investment decisions, the information ratio involves a ratio of two quantities where the numerator is the return, potentially decomposable into independent returns. Yet, the denominator (representing the standard deviation of daily returns) adds a non-linearity and necessitates modeling multivariate returns directly. Future work may explore a multivariate version of AdaVol to enhance the adaptability of our strategy in the investment decision task.

\paragraph{Optimization challenges}
Our last interpretation is technical. It stems from the convex nature of the RPS metric in contrast to the non-convex nature of the negative information ratio. Our optimization procedure is a gradient descent; it easily identifies optimal points for the probabilistic forecasting task, while this is not guaranteed for the investment decisions task. This observation highlights a potential avenue for further refinement in our optimization approach to address the unique challenges posed by the investment decision task. In particular, an in-depth analysis of the information ratio would certainly have improved our results; an illustration is that it is better to rescale the allocation to be as small as possible (summing to 0.25 in the competition) \citep[Section 2.2.2]{stanvek2023note}.

\subsection{Future work}
Future work should delve into refining our strategy by exploring a multivariate version of AdaVol, offering a more comprehensive modeling approach for investment decisions. Indeed, incorporating correlations into AdaVol could improve stability and robustness towards ill-conditioned settings \citep{godichon2023adaptive}. Additionally, optimization procedures tailored for non-convex metrics could provide valuable insights, potentially unlocking further potential in addressing the intricacies of the investment decision task. Furthermore, running parallel versions of AdaVol with different risk-appetite combined with expert aggregation could increasing robustness \citep{wintenberger2017optimal}; indeed, it should be noted that expert aggregation has provided very competitive results in various competitions \citep{gaillard2016additive, de2022state}. 

At last, Bayesian algorithm should be tested to adapt the GARCH coefficients in the same setting as AdaVol; it has been shown that state-space models yield a good representation to adapt machine learning models \cite{de2022state}. We believe this framework could be applied to the adaptation of GARCH.

In conclusion, the success in the forecasting task, coupled with the identified challenges in investment decisions, motivates us to continue refining and expanding our approach. Continuous exploration and adaptation will be crucial in unlocking the full potential of our adaptive volatility method in the realm of online probabilistic forecasting and investment decision-making.

\section*{Acknowledgement}
J. de Vilmarest and N. Werge wish to express their gratitude to Olivier Wintenberger for his invaluable support in the M6 competition.
N. Werge acknowledges the support of the Novo Nordisk Foundation (NNF) through grant number NNF21OC0070621.




\bibliographystyle{elsarticle-harv}
\biboptions{sort&compress}
\bibliography{references}

\begin{thebibliography}{30}
\expandafter\ifx\csname natexlab\endcsname\relax\def\natexlab#1{#1}\fi
\providecommand{\url}[1]{\texttt{#1}}
\providecommand{\href}[2]{#2}
\providecommand{\path}[1]{#1}
\providecommand{\DOIprefix}{doi:}
\providecommand{\ArXivprefix}{arXiv:}
\providecommand{\URLprefix}{URL: }
\providecommand{\Pubmedprefix}{pmid:}
\providecommand{\doi}[1]{\href{http://dx.doi.org/#1}{\path{#1}}}
\providecommand{\Pubmed}[1]{\href{pmid:#1}{\path{#1}}}
\providecommand{\bibinfo}[2]{#2}
\ifx\xfnm\relax \def\xfnm[#1]{\unskip,\space#1}\fi
\bibitem[{Berkes et~al.(2003)Berkes, Horv\'ath and Kokoszka}]{berkes2003}
\bibinfo{author}{Berkes, I.}, \bibinfo{author}{Horv\'ath, L.},
  \bibinfo{author}{Kokoszka, P.}, \bibinfo{year}{2003}.
\newblock \bibinfo{title}{{GARCH} processes: structure and estimation}.
\newblock \bibinfo{journal}{Bernoulli} \bibinfo{volume}{9(2)},
  \bibinfo{pages}{201--227}.
\bibitem[{Bollerslev(1986)}]{bollerslev1986generalized}
\bibinfo{author}{Bollerslev, T.}, \bibinfo{year}{1986}.
\newblock \bibinfo{title}{Generalized autoregressive conditional
  heteroskedasticity}.
\newblock \bibinfo{journal}{Journal of Econometrics} \bibinfo{volume}{31},
  \bibinfo{pages}{307--327}.
\bibitem[{Bottou et~al.(2018)Bottou, Curtis and
  Nocedal}]{bottou2018optimization}
\bibinfo{author}{Bottou, L.}, \bibinfo{author}{Curtis, F.E.},
  \bibinfo{author}{Nocedal, J.}, \bibinfo{year}{2018}.
\newblock \bibinfo{title}{Optimization methods for large-scale machine
  learning}.
\newblock \bibinfo{journal}{SIAM Review} \bibinfo{volume}{60},
  \bibinfo{pages}{223--311}.
\bibitem[{Box et~al.(2015)Box, Jenkins, Reinsel and Ljung}]{box2015time}
\bibinfo{author}{Box, G.E.}, \bibinfo{author}{Jenkins, G.M.},
  \bibinfo{author}{Reinsel, G.C.}, \bibinfo{author}{Ljung, G.M.},
  \bibinfo{year}{2015}.
\newblock \bibinfo{title}{Time series analysis: forecasting and control}.
\newblock \bibinfo{publisher}{John Wiley \& Sons}.
\bibitem[{Brockwell and Davis(2002)}]{brockwell2002introduction}
\bibinfo{author}{Brockwell, P.J.}, \bibinfo{author}{Davis, R.A.},
  \bibinfo{year}{2002}.
\newblock \bibinfo{title}{Introduction to time series and forecasting}.
\newblock \bibinfo{publisher}{Springer}.
\bibitem[{Cesa-Bianchi and Lugosi(2006)}]{cesa2006prediction}
\bibinfo{author}{Cesa-Bianchi, N.}, \bibinfo{author}{Lugosi, G.},
  \bibinfo{year}{2006}.
\newblock \bibinfo{title}{Prediction, learning, and games}.
\newblock \bibinfo{publisher}{Cambridge University Press}.
\bibitem[{Duchi et~al.(2011)Duchi, Hazan and Singer}]{duchi2011adaptive}
\bibinfo{author}{Duchi, J.}, \bibinfo{author}{Hazan, E.},
  \bibinfo{author}{Singer, Y.}, \bibinfo{year}{2011}.
\newblock \bibinfo{title}{Adaptive subgradient methods for online learning and
  stochastic optimization.}
\newblock \bibinfo{journal}{Journal of Machine Learning Research}
  \bibinfo{volume}{12}.
\bibitem[{Engle(1982)}]{engle1982autoregressive}
\bibinfo{author}{Engle, R.F.}, \bibinfo{year}{1982}.
\newblock \bibinfo{title}{Autoregressive conditional heteroscedasticity with
  estimates of the variance of united kingdom inflation}.
\newblock \bibinfo{journal}{Econometrica: Journal of the econometric society} ,
  \bibinfo{pages}{987--1007}.
\bibitem[{Francq et~al.(2011)Francq, Horvath and
  Zako{\"\i}an}]{francq2011merits}
\bibinfo{author}{Francq, C.}, \bibinfo{author}{Horvath, L.},
  \bibinfo{author}{Zako{\"\i}an, J.M.}, \bibinfo{year}{2011}.
\newblock \bibinfo{title}{Merits and drawbacks of variance targeting in garch
  models}.
\newblock \bibinfo{journal}{Journal of Financial Econometrics}
  \bibinfo{volume}{9}, \bibinfo{pages}{619--656}.
\bibitem[{Francq and Zakoïan(2004)}]{francq2004}
\bibinfo{author}{Francq, C.}, \bibinfo{author}{Zakoïan, J.M.},
  \bibinfo{year}{2004}.
\newblock \bibinfo{title}{Maximum likelihood estimation of pure garch and
  arma-garch processes}.
\newblock \bibinfo{journal}{Bernoulli} \bibinfo{volume}{10},
  \bibinfo{pages}{605--637}.
\bibitem[{Gaillard et~al.(2016)Gaillard, Goude and
  Nedellec}]{gaillard2016additive}
\bibinfo{author}{Gaillard, P.}, \bibinfo{author}{Goude, Y.},
  \bibinfo{author}{Nedellec, R.}, \bibinfo{year}{2016}.
\newblock \bibinfo{title}{Additive models and robust aggregation for gefcom2014
  probabilistic electric load and electricity price forecasting}.
\newblock \bibinfo{journal}{International Journal of Forecasting}
  \bibinfo{volume}{32}, \bibinfo{pages}{1038--1050}.
\bibitem[{Godichon-Baggioni and Werge(2023)}]{godichon2023adaptive}
\bibinfo{author}{Godichon-Baggioni, A.}, \bibinfo{author}{Werge, N.},
  \bibinfo{year}{2023}.
\newblock \bibinfo{title}{On adaptive stochastic optimization for streaming
  data: A {N}ewton's method with {O}(dn) operations}.
\newblock \bibinfo{journal}{arXiv preprint arXiv:2311.17753} .
\bibitem[{Godichon-Baggioni et~al.(2023a)Godichon-Baggioni, Werge and
  Wintenberger}]{godichon2023learning}
\bibinfo{author}{Godichon-Baggioni, A.}, \bibinfo{author}{Werge, N.},
  \bibinfo{author}{Wintenberger, O.}, \bibinfo{year}{2023}a.
\newblock \bibinfo{title}{Learning from time-dependent streaming data with
  online stochastic algorithms}.
\newblock \bibinfo{journal}{Transactions on Machine Learning Research} .
\bibitem[{Godichon-Baggioni et~al.(2023b)Godichon-Baggioni, Werge and
  Wintenberger}]{godichon2023non}
\bibinfo{author}{Godichon-Baggioni, A.}, \bibinfo{author}{Werge, N.},
  \bibinfo{author}{Wintenberger, O.}, \bibinfo{year}{2023}b.
\newblock \bibinfo{title}{Non-asymptotic analysis of stochastic approximation
  algorithms for streaming data}.
\newblock \bibinfo{journal}{ESAIM: Probability and Statistics}
  \bibinfo{volume}{27}, \bibinfo{pages}{482--514}.
\bibitem[{Hamilton(2020)}]{hamilton2020time}
\bibinfo{author}{Hamilton, J.D.}, \bibinfo{year}{2020}.
\newblock \bibinfo{title}{Time series analysis}.
\newblock \bibinfo{publisher}{Princeton University Press}.
\bibitem[{Hyndman and Athanasopoulos(2018)}]{hyndman2018forecasting}
\bibinfo{author}{Hyndman, R.J.}, \bibinfo{author}{Athanasopoulos, G.},
  \bibinfo{year}{2018}.
\newblock \bibinfo{title}{Forecasting: principles and practice}.
\newblock \bibinfo{publisher}{OTexts}.
\bibitem[{Kingma and Ba(2015)}]{kingma2015adam}
\bibinfo{author}{Kingma, D.P.}, \bibinfo{author}{Ba, J.}, \bibinfo{year}{2015}.
\newblock \bibinfo{title}{Adam: A method for stochastic optimization}, in:
  \bibinfo{booktitle}{International Conference on Learning Representations}.
\bibitem[{Makridakis et~al.(1982)Makridakis, Andersen, Carbone, Fildes, Hibon,
  Lewandowski, Newton, Parzen and Winkler}]{makridakis1982accuracy}
\bibinfo{author}{Makridakis, S.}, \bibinfo{author}{Andersen, A.},
  \bibinfo{author}{Carbone, R.}, \bibinfo{author}{Fildes, R.},
  \bibinfo{author}{Hibon, M.}, \bibinfo{author}{Lewandowski, R.},
  \bibinfo{author}{Newton, J.}, \bibinfo{author}{Parzen, E.},
  \bibinfo{author}{Winkler, R.}, \bibinfo{year}{1982}.
\newblock \bibinfo{title}{The accuracy of extrapolation (time series) methods:
  Results of a forecasting competition}.
\newblock \bibinfo{journal}{Journal of Forecasting} \bibinfo{volume}{1},
  \bibinfo{pages}{111--153}.
\bibitem[{Makridakis et~al.(1993)Makridakis, Chatfield, Hibon, Lawrence, Mills,
  Ord and Simmons}]{makridakis1993m2}
\bibinfo{author}{Makridakis, S.}, \bibinfo{author}{Chatfield, C.},
  \bibinfo{author}{Hibon, M.}, \bibinfo{author}{Lawrence, M.},
  \bibinfo{author}{Mills, T.}, \bibinfo{author}{Ord, K.},
  \bibinfo{author}{Simmons, L.F.}, \bibinfo{year}{1993}.
\newblock \bibinfo{title}{The {M2}-competition: A real-time judgmentally based
  forecasting study}.
\newblock \bibinfo{journal}{International Journal of forecasting}
  \bibinfo{volume}{9}, \bibinfo{pages}{5--22}.
\bibitem[{Makridakis and Hibon(2000)}]{makridakis2000m3}
\bibinfo{author}{Makridakis, S.}, \bibinfo{author}{Hibon, M.},
  \bibinfo{year}{2000}.
\newblock \bibinfo{title}{The {M3}-competition: results, conclusions and
  implications}.
\newblock \bibinfo{journal}{International Journal of Forecasting}
  \bibinfo{volume}{16}, \bibinfo{pages}{451--476}.
\bibitem[{Makridakis et~al.(2020)Makridakis, Spiliotis and
  Assimakopoulos}]{makridakis2020m4}
\bibinfo{author}{Makridakis, S.}, \bibinfo{author}{Spiliotis, E.},
  \bibinfo{author}{Assimakopoulos, V.}, \bibinfo{year}{2020}.
\newblock \bibinfo{title}{The {M4} competition: 100,000 time series and 61
  forecasting methods}.
\newblock \bibinfo{journal}{International Journal of Forecasting}
  \bibinfo{volume}{36}, \bibinfo{pages}{54--74}.
\bibitem[{Makridakis et~al.(2022a)Makridakis, Spiliotis and
  Assimakopoulos}]{makridakis2022m5acc}
\bibinfo{author}{Makridakis, S.}, \bibinfo{author}{Spiliotis, E.},
  \bibinfo{author}{Assimakopoulos, V.}, \bibinfo{year}{2022}a.
\newblock \bibinfo{title}{{M5} accuracy competition: Results, findings, and
  conclusions}.
\newblock \bibinfo{journal}{International Journal of Forecasting}
  \bibinfo{volume}{38}, \bibinfo{pages}{1346--1364}.
\bibitem[{Makridakis et~al.(2022b)Makridakis, Spiliotis and
  Assimakopoulos}]{makridakis2022m5}
\bibinfo{author}{Makridakis, S.}, \bibinfo{author}{Spiliotis, E.},
  \bibinfo{author}{Assimakopoulos, V.}, \bibinfo{year}{2022}b.
\newblock \bibinfo{title}{The {M5} competition: Background, organization, and
  implementation}.
\newblock \bibinfo{journal}{International Journal of Forecasting}
  \bibinfo{volume}{38}, \bibinfo{pages}{1325--1336}.
\bibitem[{Makridakis et~al.(2023)Makridakis, Spiliotis, Hollyman, Petropoulos,
  Swanson and Gaba}]{makridakis2023m6}
\bibinfo{author}{Makridakis, S.}, \bibinfo{author}{Spiliotis, E.},
  \bibinfo{author}{Hollyman, R.}, \bibinfo{author}{Petropoulos, F.},
  \bibinfo{author}{Swanson, N.}, \bibinfo{author}{Gaba, A.},
  \bibinfo{year}{2023}.
\newblock \bibinfo{title}{The {M6} forecasting competition: Bridging the gap
  between forecasting and investment decisions}.
\newblock \bibinfo{journal}{arXiv preprint arXiv:2310.13357} .
\bibitem[{Nocedal and Wright(1999)}]{nocedal1999numerical}
\bibinfo{author}{Nocedal, J.}, \bibinfo{author}{Wright, S.J.},
  \bibinfo{year}{1999}.
\newblock \bibinfo{title}{Numerical optimization}.
\newblock \bibinfo{publisher}{Springer}.
\bibitem[{Stan{\v{e}}k(2023)}]{stanvek2023note}
\bibinfo{author}{Stan{\v{e}}k, F.}, \bibinfo{year}{2023}.
\newblock \bibinfo{title}{A note on the m6 forecasting competition: Rank
  optimization}.
\newblock \bibinfo{journal}{Available at SSRN 4527154} .
\bibitem[{Straumann and Mikosch(2006)}]{straumannmikosch2007}
\bibinfo{author}{Straumann, D.}, \bibinfo{author}{Mikosch, T.},
  \bibinfo{year}{2006}.
\newblock \bibinfo{title}{Quasi-maximum-likelihood estimation in conditionally
  heteroscedastic time series: A stochastic recurrence equations approach}.
\newblock \bibinfo{journal}{Annals of Statistics} \bibinfo{volume}{34},
  \bibinfo{pages}{2449--2495}.
\bibitem[{de~Vilmarest and Goude(2022)}]{de2022state}
\bibinfo{author}{de~Vilmarest, J.}, \bibinfo{author}{Goude, Y.},
  \bibinfo{year}{2022}.
\newblock \bibinfo{title}{State-space models for online post-covid electricity
  load forecasting competition}.
\newblock \bibinfo{journal}{IEEE Open Access Journal of Power and Energy}
  \bibinfo{volume}{9}, \bibinfo{pages}{192--201}.
\bibitem[{Werge and Wintenberger(2022)}]{werge2022adavol}
\bibinfo{author}{Werge, N.}, \bibinfo{author}{Wintenberger, O.},
  \bibinfo{year}{2022}.
\newblock \bibinfo{title}{Adavol: An adaptive recursive volatility prediction
  method}.
\newblock \bibinfo{journal}{Econometrics and Statistics} \bibinfo{volume}{23},
  \bibinfo{pages}{19--35}.
\bibitem[{Wintenberger(2017)}]{wintenberger2017optimal}
\bibinfo{author}{Wintenberger, O.}, \bibinfo{year}{2017}.
\newblock \bibinfo{title}{Optimal learning with bernstein online aggregation}.
\newblock \bibinfo{journal}{Machine Learning} \bibinfo{volume}{106},
  \bibinfo{pages}{119--141}.

\end{thebibliography}

\end{document}